\documentclass[3p,times,twocolumn,preprint]{elsarticle}

\usepackage{graphicx}%
\usepackage{xcolor}
\usepackage{lineno}

\graphicspath{                
{figs} }

\usepackage{amsmath,amssymb,amsfonts}%
\usepackage{amsthm}%
\usepackage{wasysym}%
\usepackage{mathrsfs}%

\usepackage[hang,flushmargin]{footmisc} 

\begin{document}

\begin{frontmatter}  \date{2026.05.17 v3}

\title{ 
Radiation Resistance of Ge-doped Multi-Mode Fiber for Optical Links in Collider Experiments  }

\author[SMU] { Datao    Gong } 
\author[IPAS]{ Suen     Hou \corref{suen} }             
\author[IPAS]{ Bo-Jing Juang } 
\author[SMU] { Chonghan Liu } 
\author[SMU] { Tiankuan Liu } 
\author[NJU] { Ming   Qi } 
\author[SMU] { Jingbo Ye }
\author[NJU] { Lei    Zhang } 
\author[SMU] { Li     Zhang }

\cortext[suen]{E-mail address: suen@as.edu.tw }

\address[SMU]{
 Southern Methodist University, Dallas, TX 75205, U.S.A. }
\address[IPAS]{
 Institute of Physics, Academia Sinica, Taipei, Taiwan 11529 }
\address[NJU]{
 Nanjing University, Nanjing, Jiangsu 210093, China }

\begin{abstract}
 
The applications of optical links in collider experiments 
provide the advantage of high-speed data transmission with 
low mass fibers over distances of a few hundred meters. 
Ge-doped multi-mode fibers are evaluated for radiation 
tolerance in ionizing doses of Co-60 gamma rays. 
The Radiation-Induced Attenuation (RIA) varies significantly 
depending on doping substances and fabrication technologies. 
A type of telecom-grade fiber has demonstrated an RIA of 0.05~dB/m 
under a total ionizing dose of 300~kGy(SiO$_2$).
The dependence on dose rate is compared in the range
between 5 Gy/hr and 1.4~kGy/hr, and the annealing recovery 
is observed after the Co-60 source is shielded. 
Temperature effects are also investigated across a range 
of $-15 \,^\circ$C to room temperature.
At cold temperatures, stagnant annealing results in a substantially higher RIA 
during irradiation.
The recovery of radiation-induced defects is typically within 
a few hours, leading to similar  
RIA levels regardless of the dose rate and temperature during exposure.
Ge-doped fibers of chosen fabrication methods are capable of 
enduring high ionizing doses for use in high-energy physics experiments.
 
\end{abstract}

\end{frontmatter}

\begin{figure*}[t!] 
  \centering
    \includegraphics[width=.98\linewidth]{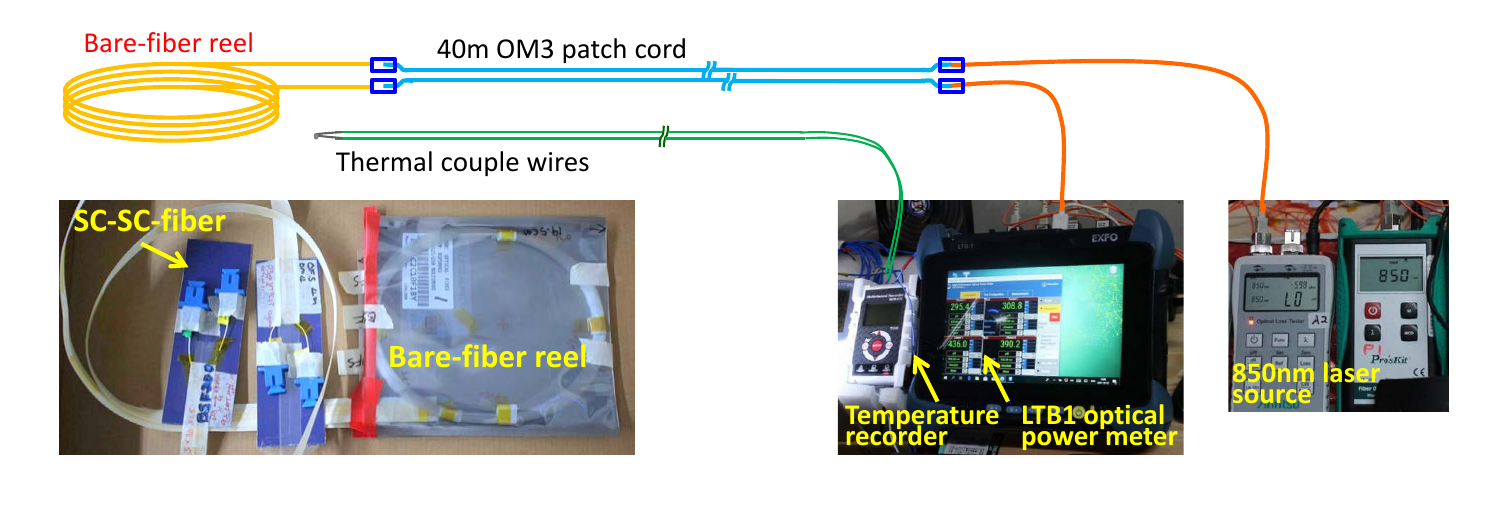}
    \vspace{-5mm}
    \caption{
    Bare-fiber reels were sealed in a water tank,
     with the fiber ends
    connected by 40~m patch cords
    to Anritsu CMA5 laser sources. The transmitted laser light
    was returned and measured by an EXPO LTB1 power meter.
    The sealed fiber samples 
    had thermal couples attached inside to record
    temperatures with an MCR-4TC.
  \label{fig:fiber-daq} }
\vspace{5mm}
\end{figure*} 

\section{Introduction }

The use of optical links in high-energy experiments provides
data transmission through low-mass fibers over distances of several hundred meters. 
The 850 nm multi-mode (MM) technology is well-suited and is 
commonly employed for data rates of 10 Gbps and higher.
The radiation tolerance of opto-electronic components under non-ionizing energy loss
has been studied for Vertical-Cavity Surface-Emitting Lasers (VCSELs)   
and photodiodes \cite{OptoNIEL-2007,OptoNIEL-2011}.
Customized transceiver ASICs have been designed to ensure both functionality 
and radiation resistance.
Recent evaluations using Co-60 gamma rays have also been reported 
\cite{MTXTID-2024}.

Optical fibers are required for radiation resistance in high-energy applications.
For instance, at the Large Hadron Collider (LHC), the inner pixel detectors 
are estimated for exposure up to 10 MGy at a rate of $\sim$200~Gy/hr \cite{LHC10Mgy}. 
The detectors are maintained at $-20 \,^\circ$C to minimize radiation damage.
The peripheral opto-links are required to endure cumulative doses of up to 1~MGy.
In comparison,
at the Circular Electron Positron Collider (CEPC), 
the expected exposure for the vertex detector is 34~kGy/y \cite{CEPCTID}.

Ge-doped optical fibers have been evaluated for radiation resistance under
ionizing radiation \cite{Oxford,RDTM2025}. 
The degradation of transmitted optical power 
through fiber is expressed by Radiation Induced Attenuation 
(RIA) versus total ionizing dose (TID):
\begin{equation}
 RIA = 10\log_{10}(\frac{P_{t=0}}{P_t})/L,
\end{equation}
where $P_t$ is the optical power transmitted through fiber length $L$, 
with the cumulative dose $t$.

High-speed telecom fibers rated for 10 Gbps (OM3, OM4) are predominantly 
of the Ge-doping type.
However, the differences in dopants and fabrication techniques, 
such as the MCVD\footnote{modified chemical vapor deposition} 
and PCVD\footnote{plasma chemical vapor deposition},  
 lead to differing performances 
under radiation.
In the following, we report the results of radiation tests 
using Co-60 gamma rays on Ge-doped multi-mode fibers
procured from various manufacturers.

The tests were conducted at the gamma facility of
INER\footnote{Institute of Nuclear Energy Research, Tauyuan, Taiwan},
which employs an array of Co-60 pellets of $\diameter 10$~mm, in a 
configuration measuring $45 \times 300$ cm$^2$.
The Co-60 array is submerged in a deep pool filled with demineralized 
water when not in use, and is raised into a shielded compartment
during irradiation service.
The irradiations of fibers were conducted at dose rates ranging from
3~Gy(SiO$_2$)/hr to 1.4 kGy(SiO$_2$)/hr by adjusting the
distances between the samples and the Co-60 array.

The cumulated ionizing doses were calibrated using Alanine dosimeters
  attached to
the fibers, which were subsequently measured with an  
 EPR analyzer\footnote{  
 Bruker EMS-104 electron paramagnetic resonance analyzer}
to determine doses with a precision of 1~\%.
For Co-60 gamma rays with energies of 1.17 MeV and 1.33 MeV,
the mass-energy absorption coefficients for Alanine and SiO$_2$ are comparable.
The dose conversion factor applied is 1~Gy(SiO$_2$) = 0.93~Gy(Alanine) \cite{Ravotti}.

The amount of radiation-induced defects in the fibers 
depends on the dose rate and the temperature of the fiber.
  The use
of fibers in high-energy experimental environments is 
considered for low temperatures and high dose rates.

\section{Fiber Co-60 irradiation setup } 

The schematics for the fiber irradiation test 
and data acquisition are 
  shown
in Fig.~\ref{fig:fiber-daq}.
Bare fiber reels were prepared with fiber ends 
terminated using 2.5~mm ferrules for connection to 
laser light sources and power meters. 
Fiber lengths of up to 1~km were optimized for approximately 10~\%
attenuation during initial hours of irradiation. 
The sealed fiber samples were immersed in a water tank,
with temperatures regulated by a metal plate connected to  
 an
external water bath. 
For temperatures below $0\,^\circ$C, the tank 
was chilled by an evaporator plate of a fridge compressor. 
Temperature monitoring was conducted using thermocouples  
 attached
to the system.

Once the fiber samples were positioned for testing, the optical powers transmitted
were recorded every minute during irradiation and consistently during
off-work hours when the Co-60 source was shielded.
The data acquisition on samples typically lasted a few weeks without interruption.
Systematic uncertainties in the optical power measurements included 
the laser source variability (specified for $\pm0.15$~dB) 
and the power meter accuracy ($\pm5$~\%).
The overall error in the optical power, verified using non-irradiated fibers, 
is estimated to be 8~\%.

\begin{figure}[b!] 
  \centering
    \vspace{-4mm}
    \includegraphics[width=.85\linewidth]{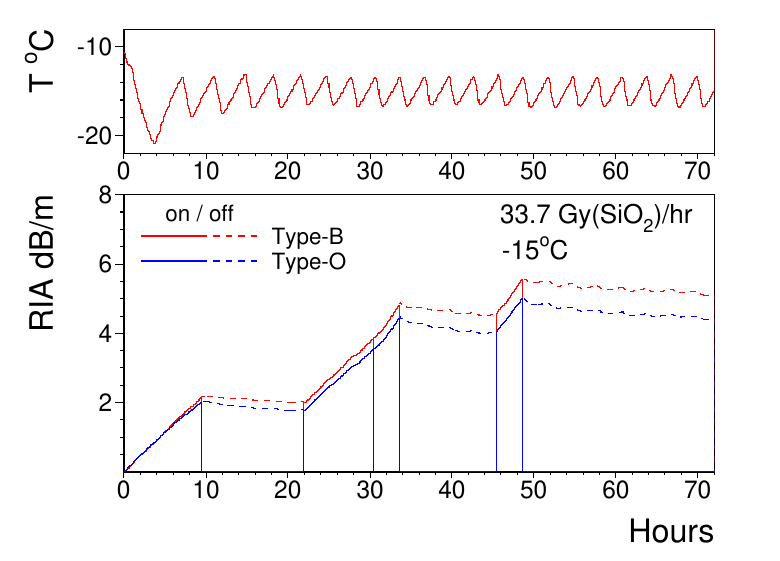}
    \vspace{-4mm}
    \caption{ 
    Two types of Ge-doped fibers (Type-B, O) showed
    significant transmission losses when exposed to ionizing radiation.
    The Co-60 irradiation tests were conducted at a dose 
    rate of 33.7 Gy/hr, at $-15 \,^\circ$C. 
    The fiber samples were immersed in a compressor-chilled water tank, 
    where the temperature fluctuated by $\pm 2\,^\circ$C due to 
    periodic switching of the control latch.
    The RIAs measured during irradiation 
    (solid line) and during subsequent annealing with the Co-60 shielded 
    (dashed line) are plotted.
  \label{fig:nonhard-doserate} }
  \vspace{-2mm}
\end{figure} 

\begin{figure}[b!] 
  \centering
    \includegraphics[width=.82\linewidth]{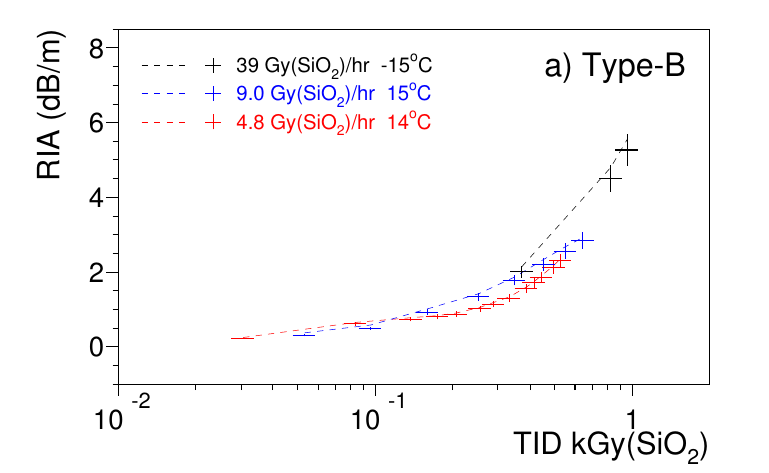}
    \includegraphics[width=.82\linewidth]{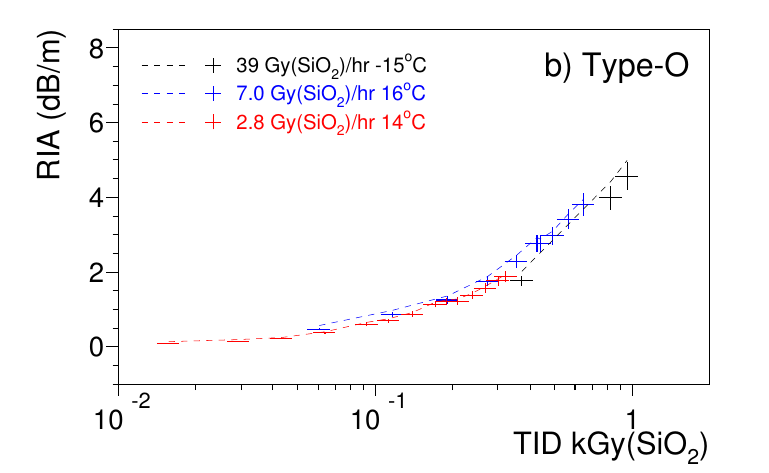}
    \vspace{-4mm}
    \caption{ Samples of two types of non-radiation hard fibers were tested
    at dose rates ranging from 2.8 to 39 Gy/hr, and at temperatures of $-15\,^\circ$C and 
    $15\,^\circ$C.
    The dashed lines represent the instantaneous RIAs at daily accumulated doses,
    while the markers indicate the corresponding RIAs after 10 hours of annealing. 
    The RIA increased linearly with TID, showing little annealing recovery.
    The overlap of data points from different samples 
    suggests that there is little dependence on dose rate and temperature.
  \label{fig:nonhard-ria} }
\end{figure} 

\section{ Characteristics of non-radhard Ge-doped fibers  }  
\label{sec:nonHard}

The formation of radiation-induced defects in optical fibers  
depends significantly on the fiber fabrication technologies and the dopants used. 
The defects, once formed, may not be easily recovered. 

Two types of fibers tested exhibited 
linear degradation under ionizing radiation.
Fig.~\ref{fig:nonhard-doserate} shows the RIA measurements
over the initial three days of irradiation.
The solid lines represent data collected during irradiation, 
while the dashed lines are those measured when the Co-60 source
was shielded during non-operational hours. 
The applied dose rate was 33.7~Gy/hr. 
The samples were maintained at $-15 \,^\circ$C with a compressor cooling plate.
The temperature fluctuations between $-13 \,^\circ$C and $-17 \,^\circ$C 
occurred due to the relay cycling,   resulting in the 
 observed zagged pattern.
The annealing gains a slight recovery of optical power, totaling less than 5~\%.  

Samples tested at lower dose rates and room temperatures are compared
and plotted in Fig.~\ref{fig:nonhard-ria}. 
The dashed lines represent RIAs versus the daily recorded TIDs, 
while the markers indicate the RIAs observed after 10 hours of annealing. 
Despite the variations in dose rates and temperatures,
the data show substantial overlap in RIA values across the samples.
The RIAs reached as high as 4~dB/m at a cumulative dose of 1 kGy(SiO$_2$). 
These findings suggest that the radiation-induced defects
increase with TID, showing minimal dependency on dose rate and temperature.

\begin{figure}[b!] 
  \centering
    \vspace{-2mm}
    \includegraphics[width=.90\linewidth]{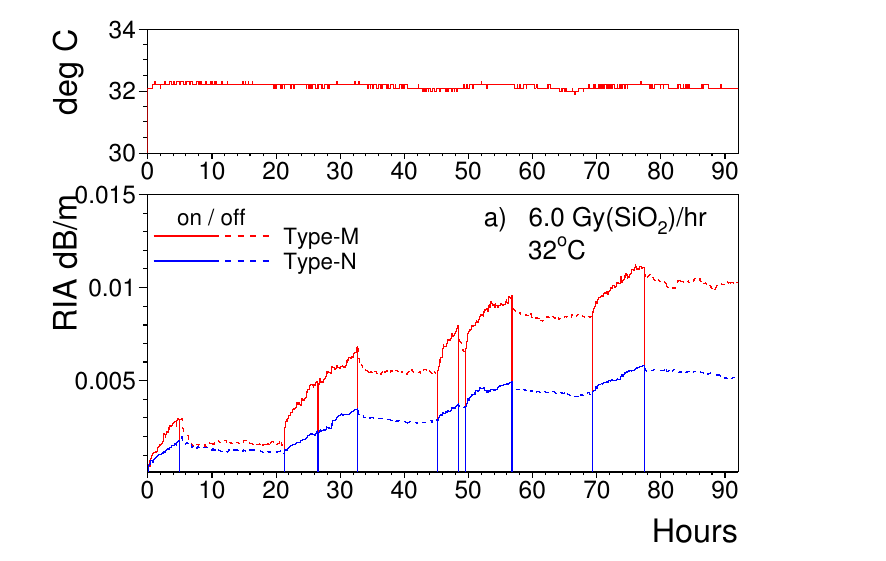}
    \includegraphics[width=.90\linewidth]{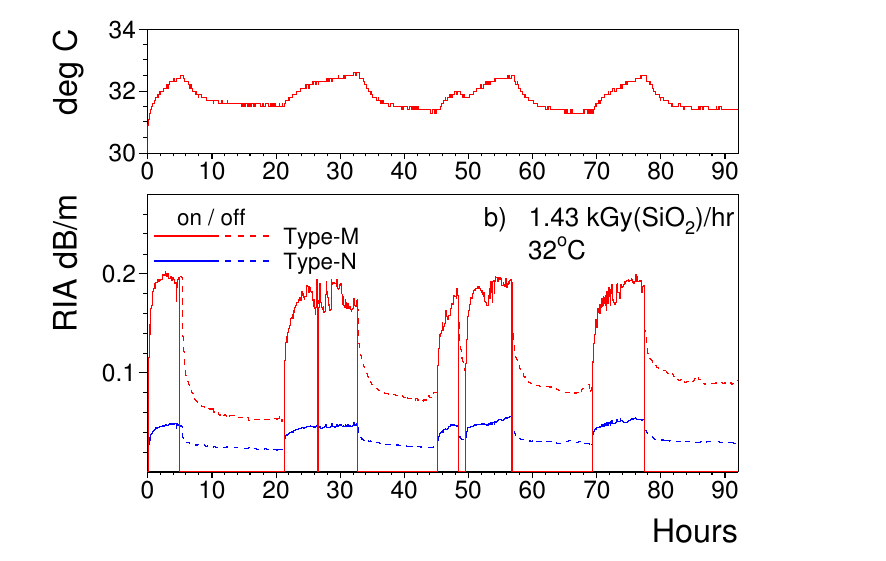}
    \vspace{-4mm}
    \caption{The RIA measurements are plotted for two types of 
    radiation-resistant 
    Ge-doped fibers during the initial days at dose rates of a) 
    6 Gy/hr and b) 1.43 kGy/hr, at 32 $^\circ$C.
    The solid lines represent data recorded during irradiation, 
    and the dashed lines indicate the measurements taken during  
    annealing with Co-60 shielded.
    At the dose rate of 1.43 kGy/hr, the radiation increased the temperature
    of the water
    tank by $1\,^\circ$C. 
    The instant RIAs are twice as high as those measured after annealing 
    for both types of fibers.
  \label{fig:radhard-doserate} }
\vspace{5mm}
\end{figure} 

\section{Radiation resistant Ge-doped fibers }  
\label{sec:Hard}

Two types of fibers exhibit strong resistance to radiation and can 
efficiently recover from ionizing defects. 
The recovery process was analyzed at different dose rates and temperatures.

At low dose rates ($\lesssim$100 Gy/hr) and room temperature, 
the annealing recovery proves to be effective. 
Fig.~\ref{fig:radhard-doserate}.a presents data collected at a dose rate of  
6.0~Gy(SiO$_2$)/hr and a temperature of 32~$^\circ$C.
The RIAs were increasing during the initial four days  (solid lines), 
with annealing recovery of less than 30~\% (dashed lines) 
after the Co-60 source was shielded.
Overall, the annealing process is effective, 
achieving substantial recovery within two hours.

\begin{figure}[b!] 
  \centering
    \vspace{-4mm}
    \includegraphics[width=1.\linewidth]{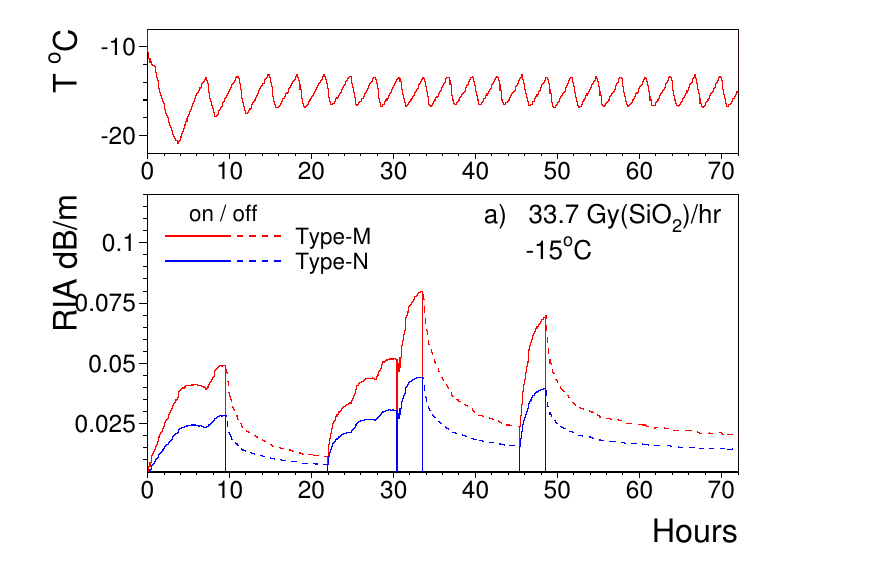}
    \includegraphics[width=1.\linewidth]{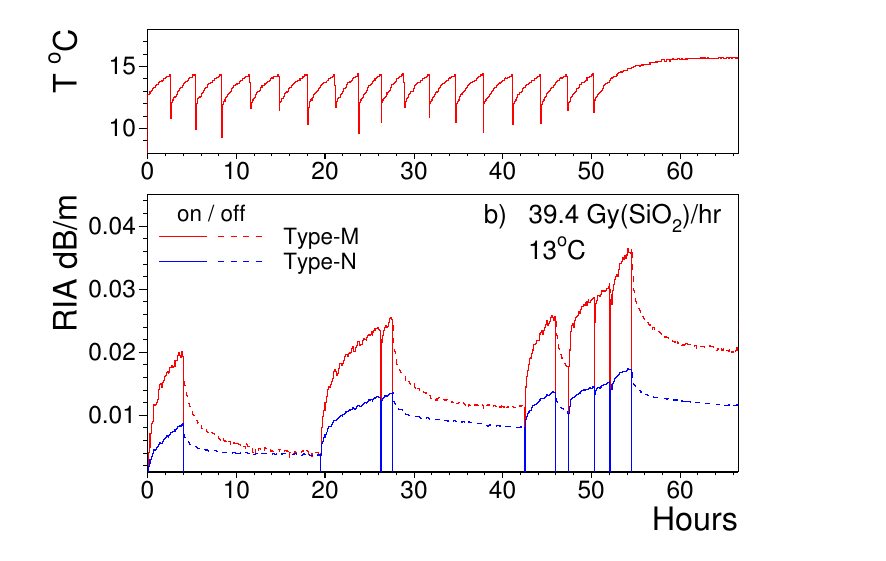}
    \vspace{-8mm}
    \caption{The irradiation tests, conducted at a dose rate of around 30 Gy/hr, are 
    compatible with the conditions present at the LHC.
     RIA tests were carried out on two radiation-resistant fibers  at  
    a) $-15 \,^\circ$C and b) $13 \,^\circ$C. The measurements from the initial 
    days are presented, showing periods of irradiation (solid lines)
    and annealing with the Co-60 source shielded (dashed lines).
    The samples immersed in water tanks were chilled by a compressor.
    The temperatures fluctuated due to the periodic switching of the control latch.
    At the cold temperature of $-15 \,^\circ$C, the RIAs in radiation varied with temperature,
    and the stagnant recovery resulted in RIA levels that are 
    twice as high as those at $13 \,^\circ$C.
  \label{fig:radhard-temp} }
\end{figure} 

When exposed to very high dose rates at room temperature, the ionizing 
defects accumulate extensively.
The RIA measurements conducted at 1.43~kGy/hr 
are plotted in Fig.~\ref{fig:radhard-doserate}.b,
showing the initial data recorded over four days. 
During the irradiation process, the samples were heated up by about $1 \,^\circ$C.
The instant RIAs (solid lines) were found to be twice 
as high as the levels recorded after annealing.

The irradiation test at a dose rate of 33.7~Gy/hr at $-15 \,^\circ$C replicates
the conditions present at the LHC.
The RIA measurements of the fibers are plotted in Fig.~\ref{fig:radhard-temp}.a,
while comparative measurements taken at  
  $13 \,^\circ$C
are shown in Fig.~\ref{fig:radhard-temp}.b. 
The temperatures were regulated using a compressor evaporator plate,
with fluctuations of $\pm2 \,^\circ$C due to relay switching.
The effect is evident at $-15 \,^\circ$C, where the RIAs increase with temperature.
In contrast, the dependence is less pronounced at $13 \,^\circ$C (Fig~\ref{fig:radhard-temp}.b).

Radiation-induced defects generated at cold temperatures tend to be stagnant in recovery.
At $-15 \,^\circ$C, the RIAs are accumulated to nearly double those observed at $13 \,^\circ$C.
However, once the Co-60 source is shielded, recovery occurs effectively 
regardless of the temperature.
Annealing at $-15 \,^\circ$C reduces the RIAs by a factor of three,
bringing them to levels that are compatible with those observed at $13 \,^\circ$C.


\begin{figure}[b!] 
  \centering
    \vspace{-4mm}
    \includegraphics[width=.85\linewidth]{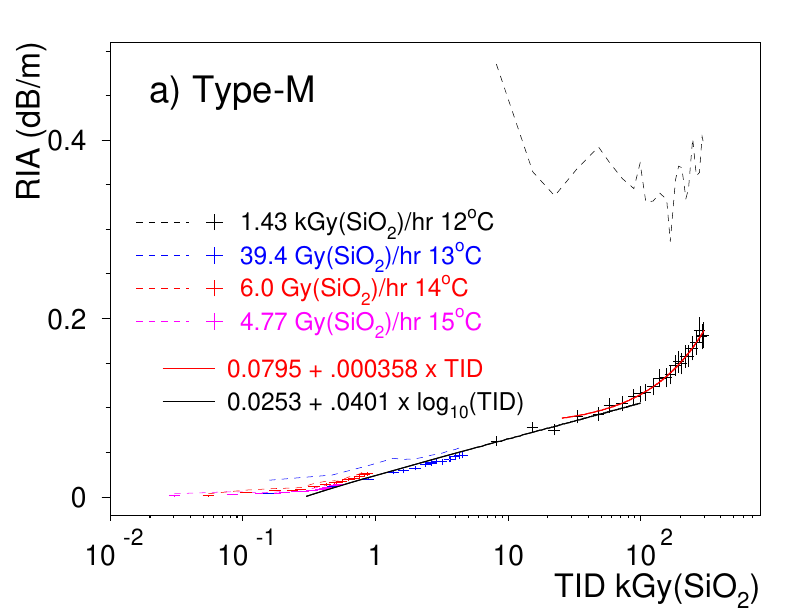} 
    \includegraphics[width=.85\linewidth]{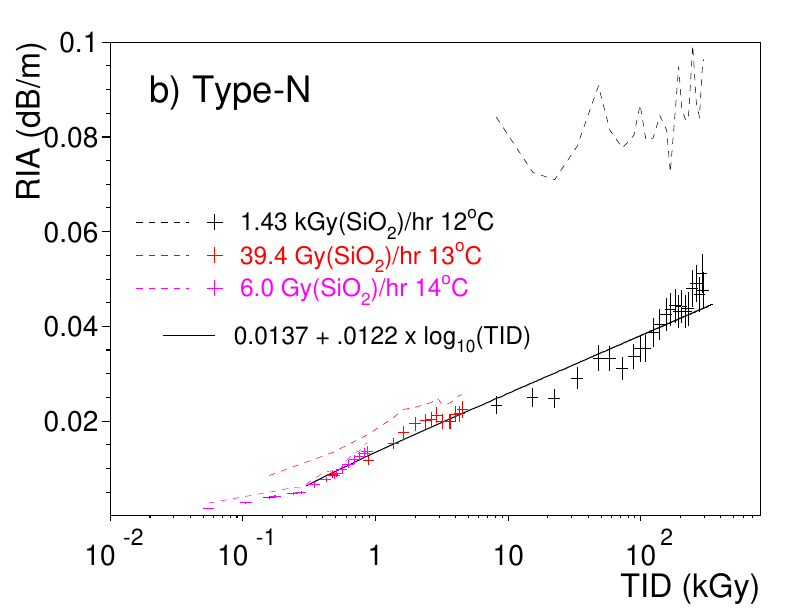} 
    \vspace{-4mm}
    \caption{
    Measurements of RIA for two radiation-resistant fiber types were taken
    at dose rates ranging from 5 to 1.43k Gy/hr around $13 \,^\circ$C.
    The dashed lines represent the instantaneous RIAs 
    at the daily accumulated doses.
    The points indicate the corresponding RIAs after 10 hours of annealing 
    with the Co-60 shielded. 
    The overlaps of annealed RIAs across samples suggests negligible
    dependency on the dose rate.
    The black lines represent fits to a logarithmic function.
    In a), the steeper RIA distribution with dose $>100$ kGy is 
    fitted using a linear function (red line).
  \label{fig:radhard-ria} }
\end{figure} 
 
To compare the dependency on dose rate,
the RIA measurements performed at 13$^\circ$C are 
  summarized
in Fig.~\ref{fig:radhard-ria}.a and b, 
for the two radiation-resistant fibers.   
  The dashed lines indicate that the instant RIAs 
  are significantly higher at a dose rate of
  1.43 kGy/hr.
The data points represent the RIAs measured after 10 hours of annealing.
Despite the wide range of irradiation dose rates, the measurements 
demonstrate consistent 
RIA levels versus the cumulative doses.

The RIA of radiation-resistant fibers follows an approximately logarithmic function of TID. 
The RIA measurements were fitted to the functions of
\begin{align}
  & RIA = a+b \cdot 10\log_{10}(TID), \label{eq:logarith} \\
  & RIA = a+b \cdot (TID).               \label{eq:linear}      
\end{align}
The fits to Eq.~\ref{eq:logarith} are represented by the black lines in 
Fig.~\ref{fig:radhard-ria}.a and b. 
In the higher TID region of $>100$~kGy, the Type-M fiber shows a steeper increase in RIA,
which is better represented by a linear function (Eq.~\ref{eq:linear})
indicated by the fitted red line in Fig.~\ref{fig:radhard-ria}.a.
The Type-N fiber demonstrates superior radiation hardness, maintaining an RIA of 0.05 dB/m 
under a cumulative dose of 300~kGy.

\section{Summary }
\label{sec:summary}

Ge-doped multi-mode fibers of telecom grades 
were evaluated for radiation tolerance to TID.
Among the tested fibers, two types exhibited strong radiation resistance.
The annealing process effectively mitigated radiation-induced defects  
within hours, with no notable dependence on dose rate or temperature.
A type of radiation-resistant fiber has demonstrated an RIA of 0.05 dB/m 
under a dose of 300 kGy(SiO$_2$).

Performance tests conducted at 34 Gy(SiO$_2$)/hr and $-15 \,^\circ$C 
align with the radiation conditions for fibers at the LHC. 
Exposure to radiation in cold temperatures can result in instant RIA 
twice as high as the annealed level.
Ge-doped fibers of chosen fabrication methods are capable of 
enduring ionizing doses for usage in high-energy experiments.


\section*{Acknowledgements}

The authors wish to extend their gratitude to the Institute of 
Nuclear Energy Research for their support. 
This work has been partially supported by the Institute of Physics, Academia Sinica.

\section*{Declarations}
{\bf Conflict of interest} On behalf of all authors, the corresponding 
author states that there is no conflict of interest.


{}

\newpage


\end{document}